\documentclass[aps, onecolumn, superscriptaddress, 12pt]{revtex4}

\usepackage[english]{babel}
\usepackage{hyperref}
\usepackage{amsfonts}

\usepackage[pdftex]{graphicx}
\usepackage{color}
\usepackage[toc,page]{appendix}

\usepackage{amsmath, bbold}
\usepackage{amssymb}
\usepackage{mathtools}
\usepackage{braket}

\usepackage[sort&compress]{natbib}

\begin{document}

\title{Revisiting the $\mathcal{PT}$-symmetric
 ring: spectra, eigenstates and transport properties}

\title{Spectra, eigenstates and transport properties of a $\mathcal{PT}$-symmetric ring}

\author{Adrian Ortega}
\email{ortega.adrian@wigner.hu}
\affiliation{Wigner  RCP,  Konkoly-Thege  M. u. 29-33, H-1121 Budapest, Hungary}

\author{Luis Benet}
\email{benet@icf.unam.mx}
\affiliation{Instituto de Ciencias F\'isicas, Universidad Nacional Aut\'onoma de M\'exico,
Av. Universidad s/n, Col. Chamilpa, C.P. 62210 Cuernavaca, Morelos, M\'exico}

\author{Hern\'an Larralde}
\email{hernan@icf.unam.mx}
\affiliation{Instituto de Ciencias F\'isicas, Universidad Nacional Aut\'onoma de M\'exico,
Av. Universidad s/n, Col. Chamilpa, C.P. 62210 Cuernavaca, Morelos, M\'exico}

\date{\today}


\begin{abstract}
We study, analytically and numerically, a simple $\mathcal{PT}$-symmetric tight-binding ring with an onsite energy $a$ at the gain and loss sites. We show that if $a\neq 0$, the system generically exhibits an unbroken $\mathcal{PT}$-symmetric phase. We study the nature of the spectrum in terms of the singularities in the complex parameter space as well as the behavior of the eigenstates at large values of the gain and loss strength. We find that in addition to the usual exceptional points, there are  ``diabolical points'', and inverse exceptional points at which complex eigenvalues reconvert into real eigenvalues. We also study the transport through the system. We calculate the total flux from the source to the drain, and how it splits along the branches of the ring. We find that while usually the density flows from the source to the drain, for certain eigenstates a stationary ``backflow'' of density from the drain to the source along one of the branches can occur. We also identify two types of singular eigenstates, i.e. states that do not depend on the strength of the gain and loss, and classify them in terms of their transport properties. 
\end{abstract}

\maketitle

\section{Introduction}
\label{sec:Introduction}

$\mathcal{PT}$-symmetry is by now a well established field of Quantum Physics, in which many interesting phenomena arise, see \cite{Bagarellobook2016, El-GanainyNature2018, Christodoulidesbook2018, Benderbook2019} and references therein. The field can be considered at a stage in which the theory is already on firm grounds and applications are starting to appear ~\cite{Chen2017,Christodoulidesbook2018,RosaJMPS2021,BergholtzRMP2021}. 
Among the many instances of systems that can be described by non-Hermitian Hamiltonians, a particularly useful one concerns their role as effective models of open systems. A simple example of such systems are tight-binding models with input (gain) and output (loss) at different sites, which are the kind of systems we are interested in this paper.

While many properties are known for a one dimensional homogeneous $\mathcal{PT}$-symmetric
tight-binding chain with open boundary conditions, see for instance ~\cite{OrtegaJPA2020, GraefeJPA2008, JinPRA2009,YogeshPRARapid2011}, the case with periodic boundary conditions has been less explored \cite{CerveroPLA2003, ZnojilPLA2011, ScottPRA2012, ScottJPA2012}. One reason for this is that, for constant values of the tunneling couplings and onsite energies, the spectrum typically becomes complex for any 
constant strength of the gain and loss
different from zero~\cite{YogeshPRARapid2011}, which motivated the study of inhomogeneous rings~\cite{ZnojilPLA2011}. 

In this work we study the spectrum and eigenvectors of a simple extension of the homogeneous $\mathcal{PT}$-symmetric tight-binding ring. We show that in spite of the periodic boundary conditions, our system generically exhibits the usual unbroken and broken $\mathcal{PT}$-symmetric phases. This is achieved by introducing complex $\mathcal{PT}$-symmetric onsite energies $\alpha=a\pm i\eta$ at two sites on the ring, the imaginary parts of which represent the gain and loss, the real part representing a real on-site energy on those sites. The rationale for this extension is the following: In the absence of the complex on-site energies, the system is rotational invariant and has multiple doubly degenerate states. In this situation, when the system is coupled to a pure gain and loss, the degenerate energy levels indeed generally, though not always, split into complex conjugate pairs and the system does not exhibit a $\mathcal{PT}$-symmetric phase. However, the inclusion of the real onsite energies (without gain and loss for the time being), retains the Hermitian nature of the system, breaks the rotational invariance, and typically lifts the degeneracies, giving place to a non-degenerate real spectrum. Since the singularities of the spectrum must occur at degeneracies, this implies that now there is a range of values of the gain and loss, from zero up until the first degeneracy occurs, for which the spectrum will continue to be real. In this range, the system is in a $\mathcal{PT}$-symmetric phase. This simple construction contrasts with previous analytical~\cite{ScottPRA2012} and numerical models~\cite{ScottJPA2012} in which one has to engineer different couplings among the sites of the chain to attain a  $\mathcal{PT}$-symmetry phase.

Our model is simple enough to be treated analytically. Specifically, following a variation of the method outlined in \cite{OrtegaJPA2020}, we obtain a relatively simple equation that determines the quasi-momentum $\theta$, in terms of which we can write closed expressions for the eigenvalues and eigenstates of the system. These results allow a detailed study of the different types of singularities that appear in the spectrum. We find that in addition to exceptional points (EPs), there are singularities that resemble ``diabolical points'' \cite{BerryWilkinson84, Berry2004} and, under certain circumstances, we also find inverse exceptional points, at which complex eigenvalues reconvert into real eigenvalues.

Our results also allow us to characterize the transport properties through the system. In particular, we show that transport is efficient in the $\mathcal{PT}$-symmetric states, meaning that the inflow and outflow of density, if any, are equal. More interestingly, we calculate how the total flux from the source to the drain splits along the branches of the ring for these $\mathcal{PT}$-symmetric states. 
We find that while usually the density flows from the source to the drain along both branches at a fixed ratio that depends on the size of the system and the distance between the leads, for other eigenstates states a stationary ``backflow'' of density from the drain to the source along one of the branches can occur.

Next, we identify singular eigenstates, i.e. states that do not depend on the strength of the gain and loss, and we classify them in terms of their transport properties. We show that one kind of singular states, that only depend on the the system size and distance between the leads, have zero amplitude at the positions of the leads. 
Thus, these states cannot couple with the gain and loss. Following ~\cite{OrtegaJPA2020} these singular eigenstates are called ``opaque'' (non conducting). A different type of singular eigenstates, which we call accidental singular states, only appear for certain configurations and specific values of the onsite energy. These states do couple with the gain and loss, and as they have efficient transport, we classify these states as ``transparent'' (or conducting).

Finally, we show an interesting feature of this system, which we call ``eigenvalue reconversion''. Usually after a coalescence of energies at an EP, these become complex and remain so as the gain and loss increases. However, in our system we observe that as the strength of the gain and loss increases, some complex eigenvalues may coalesce again at a reverse EP, and reconvert to being purely real. Thus the states corresponding to these eigenvalues recover the properties of the $\mathcal{PT}$-symmetric states, for example, they are again capable of conducting efficiently across the system. Additionally, we also show how some eigenstates vanish on one branch of the ring as the strength of the input and output increases, attaining a ``directional character''.

\section{The model and its solution}
\label{sec:ModelSolution}

We consider a Hamiltonian for a periodic tight-binding chain with $N$-sites, uniform
couplings among neighboring sites, and complex $\mathcal{PT}$-symmetric self energies at sites $k$ and $k'$. The imaginary parts of these represents a gain and a loss respectively, the real parts correspond to on-site energies. The Hamiltonian reads~\cite{Benderbook2019}

\begin{equation}
  H = \sum_{i=1}^{N-1}\Big(\ket{i}\bra{i+1} + \ket{i+1}\bra{i} \Big)
  + \ket{1}\bra{N} + \ket{N}\bra{1} + \Big(\alpha\ket{k}\bra{k} + \alpha^*\ket{k'}\bra{k'} \Big),
  \label{eq:ham}
\end{equation}

where $\alpha=a+i\eta$ and $a,\eta$ are real parameters. 

With periodic boundary conditions, the relevant coordinates are the 
distances between the positions of gain and loss along each branch of the ring, i.e. $k'-k$ and $N-(k'-k)$ (for concreteness, we shall consider
that $k'>k$), or, equivalently, $k'-k$ and $N$ . 
The periodic boundary conditions allow one to choose an appropriate $\mathcal{P}$ operator for arbitrary positions of the source and drain. 
Specifically, $\mathcal{P}$ should ``reflect'' the system along the axis that splits the ring symmetrically half way between $k$ and $k'$ (note that the representation of the parity operator can be different than in the open boundary case~\cite{OrtegaJPA2020}). The time inversion operator is defined as usual, namely $\mathcal{T}A\mathcal{T} = A^*$ where ${}^*$ represents complex conjugation and $A$ is any matrix operator~\cite{WangPTRS2013}. Then, as
expected, the Hamiltonian commutes with the operator $\mathcal{PT}$. 

Following the method of Losonczi-Yueh~\cite{Losonczi1992, Yueh2005, OrtegaJPA2020}, the eigenvectors can be obtained explicitly. From Eq.~(\ref{eq:ujapp}), if we call $\theta$ the quasi-momentum, then the $u_j$ component reads
\begin{equation}
   \begin{split}
     u_j &=\frac{1}{\sin\theta}\Big[ u_0\sin[(1-j)\theta] + 
     u_1\sin(j\theta) -\alpha u_k\sin[(j-k)\theta] \Theta(j-k-1)\\ 
     &\qquad -\alpha^* u_{k'} \sin[(j-k')\theta]\Theta(j-k'-1) \Big],
   \end{split}
  \label{eq:uj}
 \end{equation}
 
 with $\Theta(x)$ the usual Heaviside function and $1\leq j\leq N$ an integer.
The eigenvalues of the system can be written as 
\begin{equation}
  E = 2\cos\theta.
  \label{eq:E}
\end{equation}
The $u_k$ and $u_{k'}$ components
are obtained from equation (\ref{eq:uj}) by evaluating at $j=k$ and $j=k'$ respectively. Thus, all the eigenvector components can be expressed in terms of $u_0$ and $u_1$. Finally, requiring that the homogeneous system that results from evaluating (\ref{eq:uj}) at $u_0$ and $u_1$ has nontrivial solutions, implies that $\theta$ must fulfill the equation
\begin{equation}
4\, \sin^2\left(\frac{N\theta}{2}\right) + 2a \frac{\sin(N\theta)}{\sin\theta} - |\alpha|^2\frac{\sin[(k'-k)\theta]\sin[(N-k'+k)\theta]}{\sin^2\theta} = 0.
  \label{eq:theta}
\end{equation}
Details of the derivation of Eq.~(\ref{eq:theta}) can be found 
in Appendix~\ref{sec:egvalsol}, which considers a slightly more general
Hamiltonian.

For $\alpha=0$, i.e. for the Hermitian rotational invariant system, the quasi-momenta that fulfill Eq.~(\ref{eq:theta}) are $\theta_0=2\pi m/N$, where $m$ is an integer that goes from 0 to $N-1$. We note that the quasi-momenta $2\pi m/N$ and $2\pi (N-m)/N$ yield the same eigenvalue $E$ for all values of $m$. 
That is, all eigenvalues are doubly degenerate except for those associated with $\theta=0$ and, if $N$ is even, for $\theta=\pi$.

We consider the case $a=0$ using standard perturbation theory, and take $\eta$ as our perturbation parameter. First, we discuss what happens to the quasi-momenta that correspond to degenerate states, and later the non-degenerate cases $\theta=0,\pi$. The unperturbed solutions have the form $\theta_0=2\pi m/N$ with $m=1, \dots, N-1$, excluding $m= N/2$ for even $N$, and we write the perturbed solutions as $\theta=\theta_0+\delta\theta_e \eta$. Expanding Eq.~(\ref{eq:theta}) to second order in $\eta$ and equating the expression to zero, we obtain
\begin{equation}
    \Big(N^2 \delta\theta_e^2 + \frac{\sin^2[(k'-k)\theta_0]}{\sin^2\theta_0}\Big)\eta^2 = 0,
\end{equation}
which yields
\begin{equation}
\delta\theta_e = i\sin[(k'-k)\theta_0]/(N\sin\theta_0).
\label{eq:delta_e}
\end{equation}
In Eq.~(\ref{eq:delta_e}) we have chosen the positive sign of the square-root, though the negative sign would be just as good; either choice spans identically the perturbed solutions, differing only in the specific labelling of the quasi-momenta to the eigenvalues.

The resulting (perturbed) quasi-momenta no longer lead to a double degeneracy: the degenerate unperturbed energies split into a complex-conjugate pair of eigenvalues for $\eta\neq 0$; the difference between these pairs being proportional to the perturbation parameter $\eta$. Thus, for $a=0$ we have a broken $\mathcal{PT}$-symmetric phase as soon as $\eta$ is different from zero, provided that $(k'-k)\theta_0\neq r\pi$. For the case $\theta_0=0$ (or $\pi$ for even $N$), we obtain $\delta\theta_e=\sqrt{(k'-k)(N-k'+k)}/N$, which is real, thus the eigenvalues corresponding to these values of $\theta_0$ remain real for $\eta \neq 0$. 

Now we consider the case with $\eta=0$ and take $a\neq 0$ as our perturbation parameter. This is a Hermitian system and therefore the eigenvalues are real. As before, we begin with quasi-momenta of the form $\theta_0=2\pi m /N$. Expanding~(\ref{eq:theta}) to second order in $a$, we obtain
\begin{equation}
    \Big(N^2\delta\theta_a^2+\frac{2N}{\sin\theta_0}\delta\theta_a
+ \frac{\sin^2[(k'-k)\theta_0]}{\sin^2\theta_0}\Big)a^2 = 0,
\label{eq:perturb_a}
\end{equation}
which yields 
\begin{equation}
\delta\theta_a = \frac{1}{N}\left(-\frac{1}{\sin\theta_0} + \frac{\cos[(k'-k)\theta_0]}{|\sin\theta_0|}\right).
\label{eq:delta_a}
\end{equation}
As in Eq.~(\ref{eq:delta_e}), we have chosen in Eq.~(\ref{eq:delta_a}) the positive solution of the square-root. Thus, the non-zero onsite energy lifts the degeneracy of the rotational symmetric $a=0$ case.
The case $\theta_0=0$ is treated similarly, but using $\theta=\sqrt{a}\delta\theta_a$ due to the singularities arising from the denominators in Eq.~(\ref{eq:theta}). In this case, the first order expansion yields a purely imaginary correction $\delta\theta_a = i\sqrt{2/N}$, and therefore the corresponding energy is $2\cosh(\sqrt{2/N})$, which is real and larger than 2. A similar approach can be used for $\theta_0=\pi$, and yields $\delta\theta_a=\sqrt{2/N}$.

\begin{figure}
  \centering
  \includegraphics[scale=0.45]{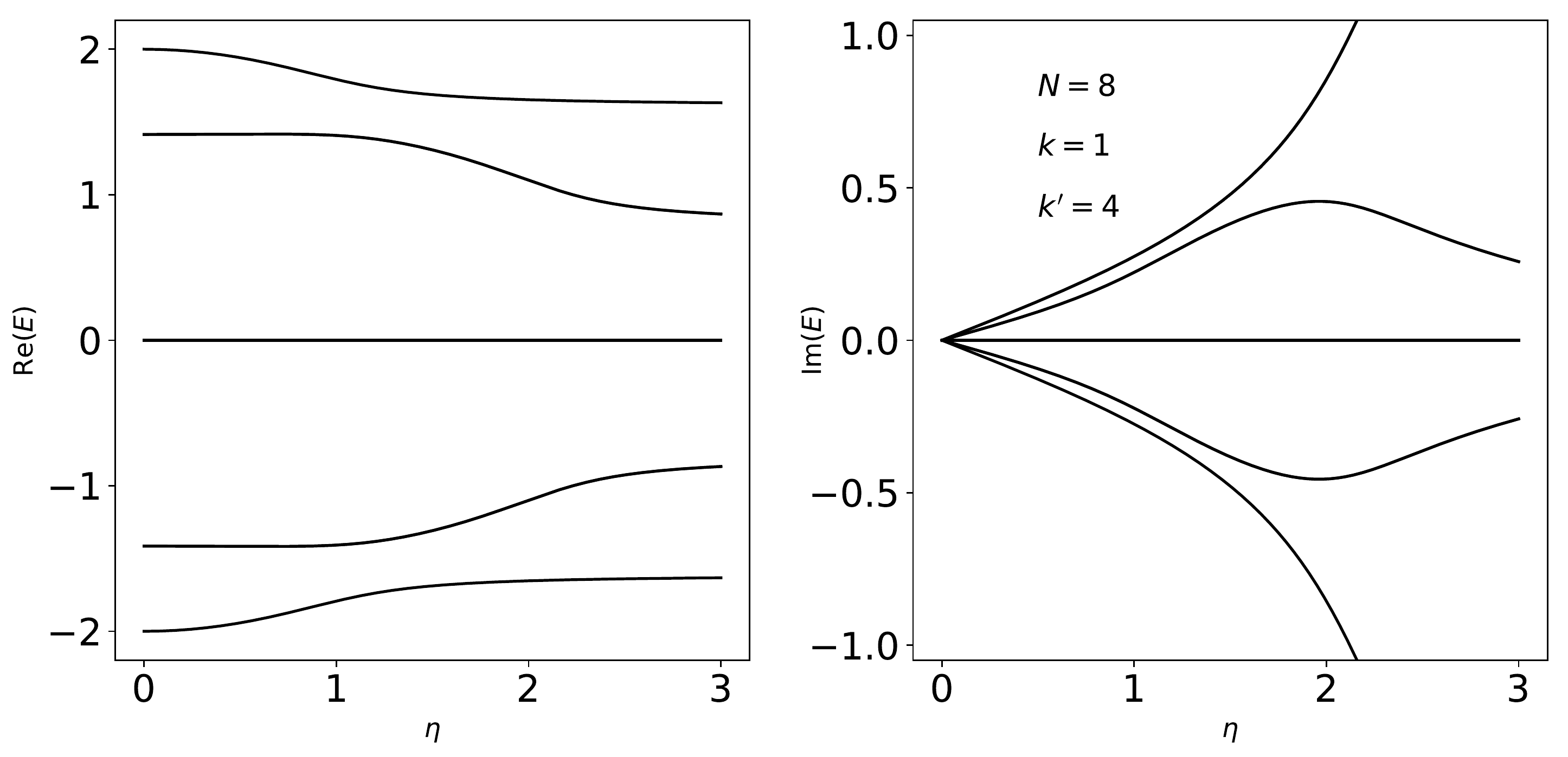}
  \includegraphics[scale=0.45]{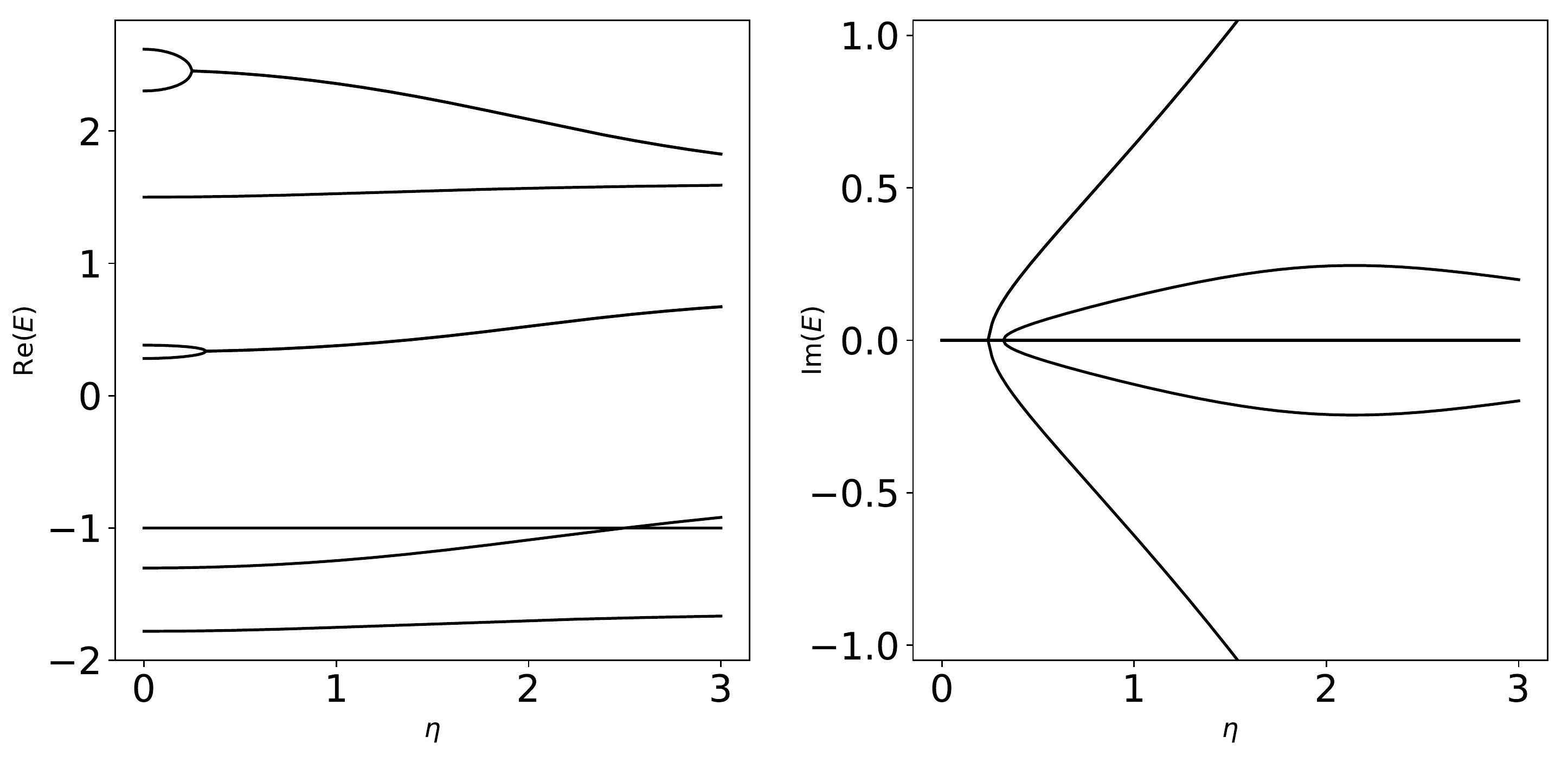}
  \caption{Spectrum for $a=0$ (top row)
  and $a=1.5$ (bottom row) as a function of $\eta$. The left panels illustrate the real part
  of the eigenvalues $E$ and the right ones their imaginary part. In the first case the spectrum is always complex for $\eta>0$ and the $\mathcal{PT}$-symmetric phase is absent. Note that the imaginary part of the eigenvalues is linear in $\pm\eta$ for $\eta$ near zero. This behavior is reminiscent of a ``diabolical'' point. If $a$ is different from zero, a $\mathcal{PT}$-symmetric phase occurs for a range of values of $\eta$.
  }
  \label{fig:uno}
\end{figure}

The central point is that, were it not for the self-energies $a$ at the leads, the presence of gain and loss would, in general, split the eigenvalues corresponding to degenerate states into complex conjugate pairs as soon as $\eta\neq 0$, and there would be no $\mathcal{PT}$-symmetric phase. The inclusion of the self-energy $a$ at the leads lifts the double degeneracies that are  present for $a=0$, and allows for a range of values of the parameter $\eta$, in which all the eigenvalues of the system are real despite the lack of Hermiticity of the Hamiltonian. That is, a $\mathcal{PT}$-symmetry phase is possible, in general, by simply introducing the self energies. These findings are illustrated in Fig.~\ref{fig:uno}. Further, as we shall see below, under certain conditions which can be guessed  from Eq.~(\ref{eq:delta_e}), there can also be a $\mathcal{PT}$-symmetric phase for $a=0$; see bottom row in Fig.~\ref{fig:singular}. 

For the case $a=0$, the point $\eta=0$ is somewhat unusual as a singularity in the complex plane, at least in comparison with the behaviour around an exceptional point. Indeed, as mentioned above, the first correction to the quasi-momenta is linear in $\eta$, which carries over to the energies. This is in contrast to the square-root singularity which is usually observed close to an exceptional point~\cite{BerryWilkinson84,Seyranianbook2003, Berry2004,SeyranianJPA2005,HeissJPA2012}. This behavior is illustrated in Fig.~\ref{fig:uno}, in the imaginary part of $E$ for $a=0$. The linear dependence is reminiscent of the so-called diabolical points, also known as conical intersections \cite{BerryWilkinson84, Berry2004}. Thus, the $\mathcal{PT}$-symmetric ring exhibits both diabolical points (at $a=0$) as well as exceptional points (when $a\neq 0)$. This property makes it an attractive system to exploit the different sensitivities that these singularities induce~\cite{Chen2017,RosaJMPS2021}.

We consider now the behaviour of the eigenvalues in the limit $\eta\rightarrow\infty$, which can also be understood using standard perturbation theory. Similarly to the case of the chain with open boundary conditions~\cite{DangelPRA2018, OrtegaJPA2020}, when $\eta\rightarrow\infty$ the system effectively separates, in general, into four subsystems: The two sites of the gain and loss, where the imaginary potential dominates, and two sub-chains with vanishing boundary conditions (unless the gain and loss are placed beside each other). Indeed, as $\eta$ becomes very large, two eigenstates become increasingly localized at the gain and at the loss, becoming decoupled from neighbouring sites. The eigenenergies of the states localized at the gain and loss are obtained by setting $\theta=i\phi$ in  Eq.~(\ref{eq:theta}). In this case, assuming $\phi\to\infty$ as $\eta$ grows, we find that it fulfills
\begin{equation}
    e^{2\phi}-2ae^{\phi}+|\alpha|^2 \to 0,
\end{equation}
whose solution is $\phi\sim\log(a\pm i\eta)$, which is consistent with our assumption. Thus, the energy of these states is given by
\begin{equation}
    E_{\theta,\eta\rightarrow\infty} = 2\cosh\phi \sim |a+i\eta|e^{i\arg (a\pm i\eta)} + \frac{1}{|a+i\eta|}e^{-i\arg(a \pm i\eta)}.
    \label{eq:Elargeeta}
\end{equation}
This expression shows that for large $\eta$, the real part of these eigenvalues tends to $a$, and the imaginary part which grows as $\pm \eta$. We notice that for $a=0$ we have $\arg(\pm i\eta)=\pi/2$ and then $E_{\theta,\eta\rightarrow\infty}\sim\pm i\left(\eta -1/\eta \right)$. This is the same result, in the corresponding limit, for the chain with open boundary conditions~\cite{OrtegaJPA2020, DangelPRA2018}.


The behaviour of the energies for the remaining values of $\theta$ is computed as follows. We first divide Eq.~(\ref{eq:theta}) by $|\alpha|^2$. In the limit $\eta\rightarrow\infty$ we use the perturbation parameter
$\epsilon=1/|\alpha|^2$. The leading order, $\theta_0$, are solutions of
\begin{equation}
 \frac{\sin[(k'-k)\theta_0]\sin[(N-k'+k)\theta_0]}{\sin^2\theta_0} = 0.
   \label{eq:accidental-states}
\end{equation}
These can be written as $\theta_0 = \pi m_1/(k'-k)$, for $m_1= 1,\dots,(k'-k)-1$, or as $\theta_0 = \pi m_2/(N-k'+k))$ for $m_2=1,\dots,(N-k'+k)-1$. To determine the nature of the first order correction in $\epsilon$, we consider the derivative of the function $g(\theta)=\sin[(k'-k)\theta]\sin[(N-k'+k)\theta]$ at $\theta_0$:
\begin{eqnarray}
  g'(\theta_0) = (k'-k)\cos[(k'-k)\theta_0]\sin[(N-k'+k)\theta_0] + \nonumber\\ \qquad (N-k'+k)\sin[(k'-k)\theta_0]\cos[(N-k'+k)\theta_0]. \label{eq:theta0}
\end{eqnarray}
If $g'(\theta_0)\neq 0$ then we can write the perturbation expansion as $\theta\approx \theta_0 + \epsilon\theta_1 +...$, where $\theta_1$ and all subsequent corrections are real. In this situation, the imaginary part of the eigenvalues is identically zero for large enough values of $\eta$, which, as we shall see, occurs either by crossing a reverse EP or if the eigenvalue was never complex. 

To analyze the case $g'(\theta_0)=0$, we write generically $\theta_0=\pi m / q$, with $m,q\in \mathbb{Z}$; then, $g'(\theta_0)=0$ occurs if $q$ divides $k'-k$ and $N$ simultaneously. If $Nm/q$ is even, $\theta_0$ is an exact solution of Eq.~(\ref{eq:theta}) independently of $\eta$ and corresponds to a ``singular state'', which we discuss in depth in Sect. \ref{sec:singular}.
If $Nm/q$ is odd, then the appropriate perturbation expansion is $\theta\approx \theta_0 + \epsilon^{1/2}\theta_1$, where
\begin{equation}
  \theta_1 = \pm 2i \frac{\sin\theta_0}{\sqrt{(k'-k)(N-k'+k)}}.
  \label{eq:theta1OddCase}
\end{equation}
In this case, a complex correction term remains and vanishes slowly as $\eta\to\infty$. Examples of these behaviors can be seen in Figs.~\ref{fig:5} and~\ref{fig:7}.

\section{Transport and eigenstate classification}

\subsection{Flux and transport}
\label{ssec:Flux}

As in the case of the linear chain~\cite{OrtegaJPA2020}, we define the local density flux as
\begin{equation}
    J_n \equiv -2\,\textrm{Im}\left( c_n(t) c_{n-1}^*(t)\right),
\end{equation}
where the coefficients $c_n(t)$ are defined by the solution to the 
time-dependent Schr\"odinger equation 
written as $|\Psi(t)\rangle = \sum_j c_j(t) |j\rangle$, in the the site basis. When the initial condition is an eigenstate of the Hamiltonian, it evolves with time dependent coefficients $c_n(t) = \exp(-i E_\theta t) u_n$. If we focus on the non-degenerate
$\mathcal{PT}$-symmetric states, then the eigenvalue $E_\theta$ is 
real, and we obtain
\begin{eqnarray}
    J_{n}(\theta) = 
    -\eta\left[1-\frac{\tan(\frac{N}{2}-k'+k)\theta}{\tan(\frac{N}{2})\theta} \right]|u_k|^2
    +2 \eta\Big(|u_k|^2\Theta(n-k-1) - |u_{k'}|^2\Theta(n-k'-1) \Big).\nonumber\\
    \label{eq:flux}
\end{eqnarray}
Since the flux must be equal for $n\leq k$ and $n>k'$, the above result also implies that $|u_k|^2=|u_{k'}|^2$ for these states.
If we denote  $J_{\rm right}$ the flux for $k<n\leq k'$, and $J_{\rm left}$ the flux for the other branch of the ring ($n\leq k$ and $n>k'$), from Eq.~(\ref{eq:flux}) we have
\begin{eqnarray}
    J_{\rm right}=2\eta\left[\frac{\sin(N-k'+k)\theta}{\sin(k'-k)\theta +\sin(N-k'+k)\theta}\right]|u_k|^2
    \label{eq.Jright}
\end{eqnarray}
and
\begin{eqnarray}
    J_{\rm left}=-2\eta\left[\frac{\sin(k'-k)\theta}{\sin(k'-k)\theta +\sin(N-k'+k)\theta}\right]|u_k|^2.
    \label{eq.Jleft}
\end{eqnarray}
Then
\begin{eqnarray}
    J_{\rm right}-J_{\rm left}=2\eta|u_k|^2,
    \label{eq:efficient_transp}
\end{eqnarray}
indicating the total transport from the source to the sink is efficient, in the sense that the inflow of density equals the total flux along the branches of the ring. 
\begin{figure}
  \centering
  \includegraphics[scale=0.5]{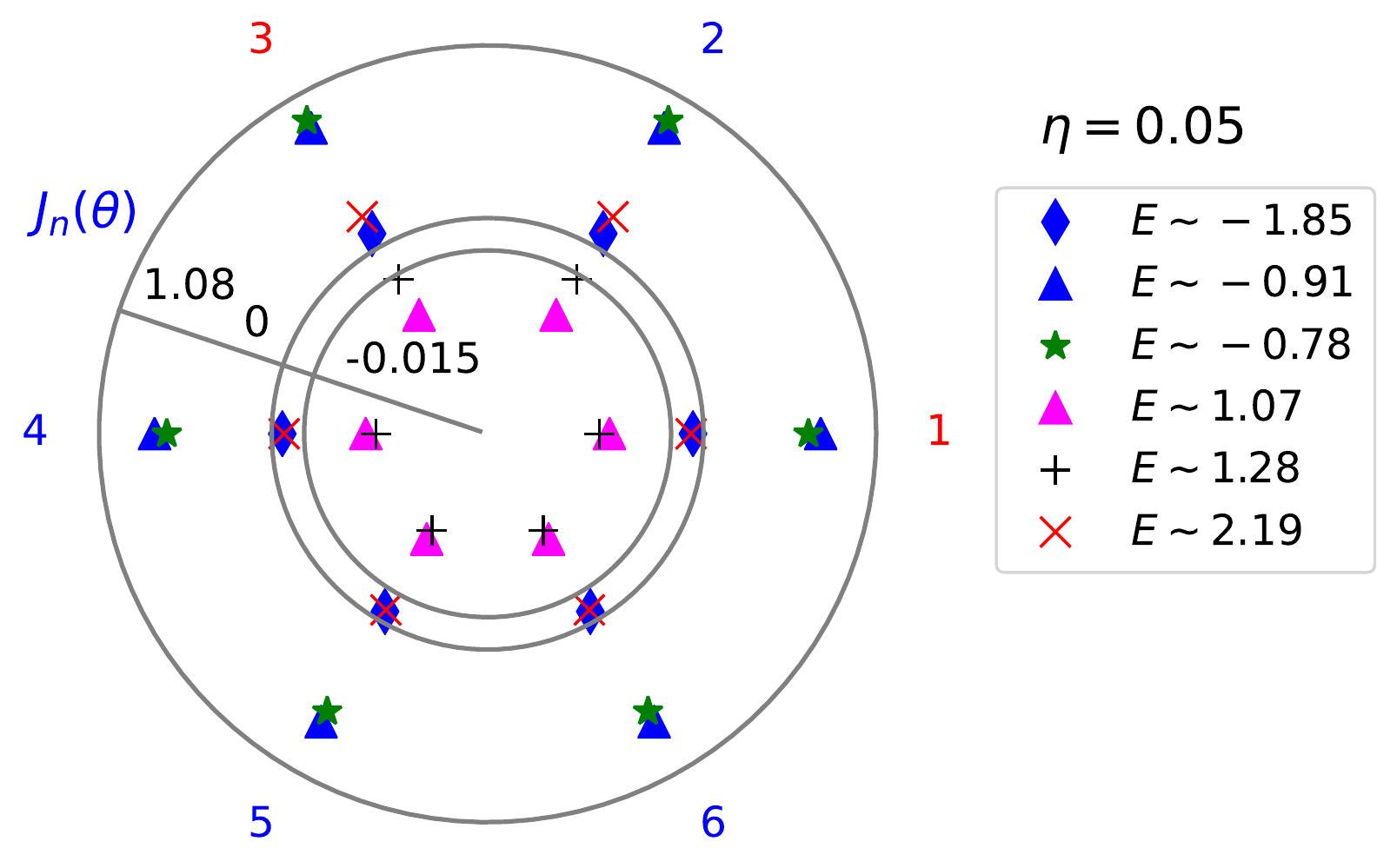}
  
  \vspace{0.5cm}
  \includegraphics[scale=0.5]{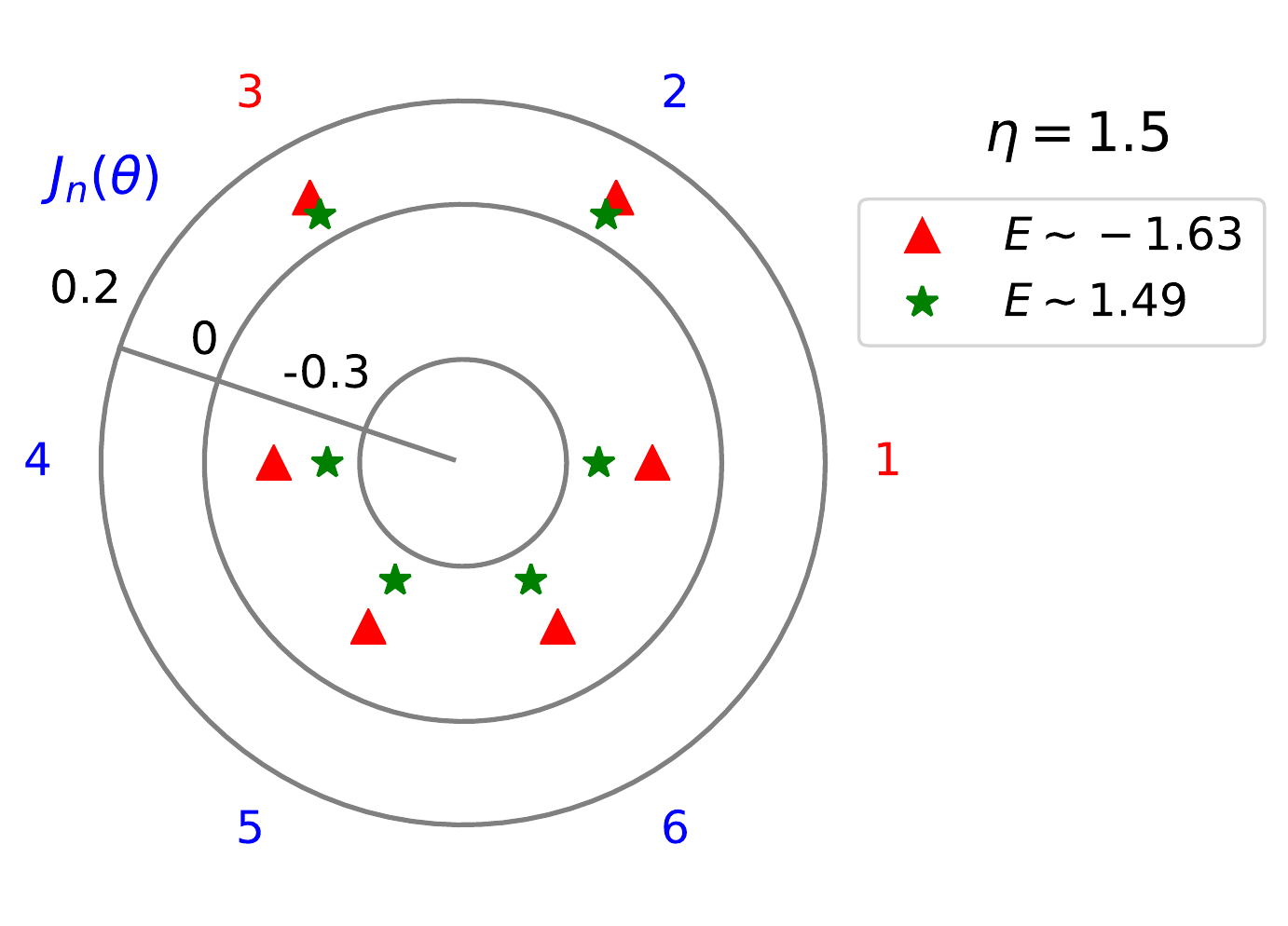}
  \includegraphics[scale=0.5]{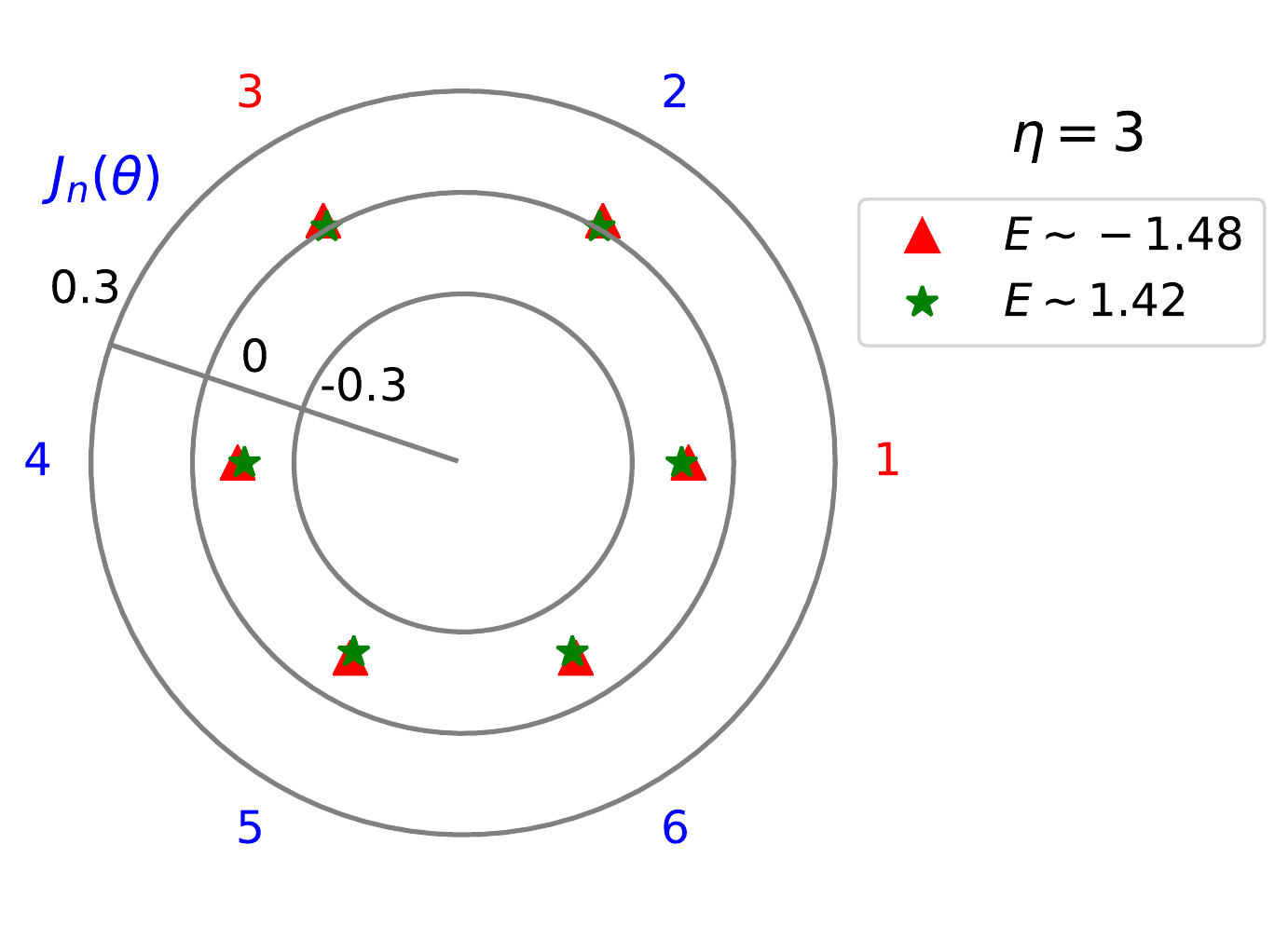}

  \caption{Stationary local fluxes $J_n(\theta)$ corresponding to purely real energies, plotted as a radial coordinate, in terms of $n$, the position along the ring for a ring, with $N=6$ and leads located at $k=1$ and $k^\prime=3$ with $a=0.5$, for different values of $\eta$. Concentric circles correspond to a constant value for the flux; for simplicity we have included three circles with the corresponding value of the flux.
  Note that when $\eta=0.05$, 
  for four eigenstates $J_{\rm right}$ and $J_{\rm left}$ have the same sign. In these cases there is a backflow from the sink to the source along one of the branches.}
  \label{fig:dos}
\end{figure}

In Fig.~\ref{fig:dos} we present the stationary local density flux $J_n(\theta)$, plotted as a radial coordinate, in terms of the position along the ring of size $N=6$, with the leads located at $k=1$ and $k^\prime=3$; we take $a=0.5$ and we consider three values of $\eta=0.05,1.5,3$. The local currents for the $\mathcal{PT}$-symmetric states are given by Eq.~(\ref{eq.Jright}) from $n=1$ through $n=3$ in the clockwise direction, and Eq.~(\ref{eq.Jleft}) in the anti-clockwise direction. Note that in the $\eta=0.05$ case, the fluxes corresponding to the states with energies $E\sim-1.85$ and $2.19$ have $J_{\rm right}>0$ and $J_{\rm left}<0$, respectively, which implies that density flows from the source to the sink on both branches. However, for all the other states, the sign of the current is the same on both branches; thus there is a back flow of density from sink to source along one branch or the other in each of these cases.  For the remaining values of $\eta$, we only show the flux for the $\mathcal{PT}$-symmetric states, the other eigenstates correspond to energies having a nonzero imaginary part, and thus, do not have stationary fluxes. 

\subsection{Singular eigenstates}
\label{sec:singular}

As in the case with open boundary conditions, eigenstates with energies that do 
not depend on the values of $\eta$ and $a$ can appear in the system. In 
\cite{OrtegaJPA2020} we called these states opaque if the wave function 
vanished at the contacts, implying that there was no transport through 
the system, 
or transparent, if the wave function did not vanish at the contacts and 
transport was efficient, in the sense, as before, that the input is equal to the output, and there is no build up nor depletion of density inside the system. 
For the time being we call 
these states ``singular states'', and later on distinguish under which 
conditions they correspond to opaque or transparent states.  

\begin{figure}
  \centering
  \includegraphics[scale=0.45]{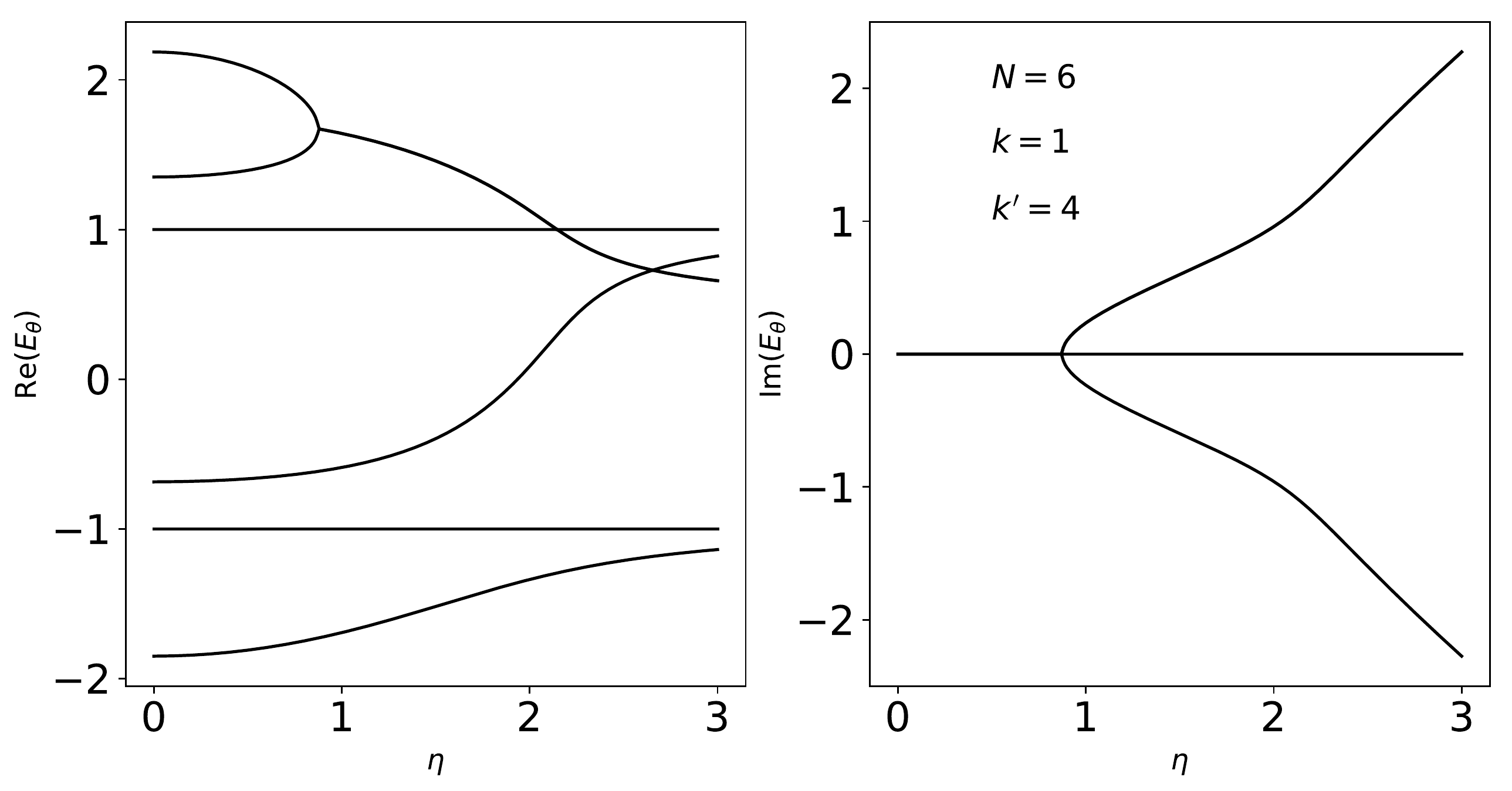}
  \includegraphics[scale=0.45]{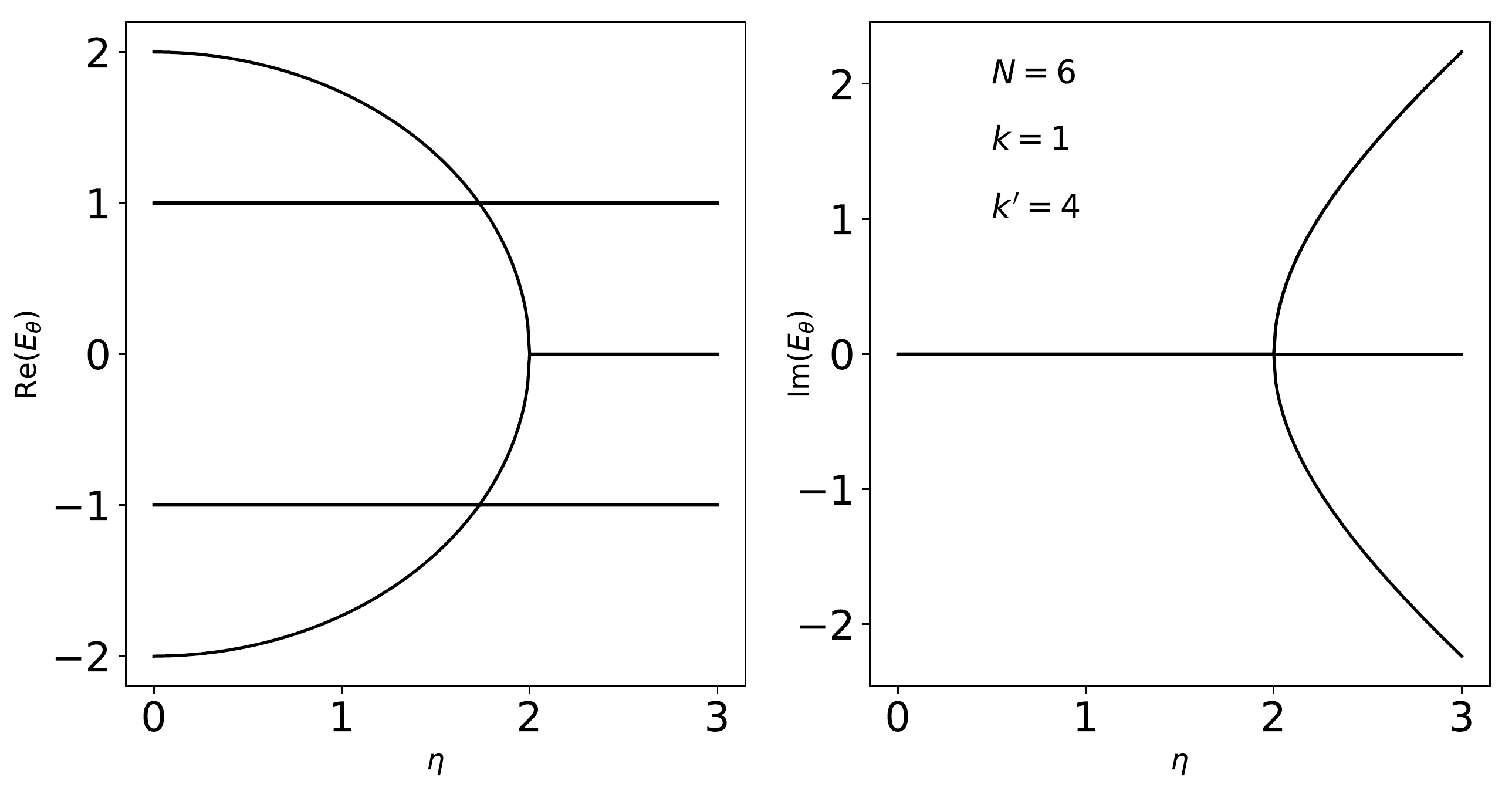}
  \caption{Spectra of a $\mathcal{PT}$-symmetric ring
  for $N=6$, $k'-k=3$, with singular states at 
  $E=\pm 1$ ($\theta_s=\pi/3, 2\pi/3$). 
  Top: $a=0.5$, the singular states are 
  non-degenerate. Bottom: $a=0$ and the singular states
  are doubly degenerate. The states that yield
  the exceptional point at $\eta=2$ satisfy
  Eq.~(\ref{eq:circle}).
  }
  \label{fig:singular}
\end{figure}

We begin by considering solutions of Eq.~(\ref{eq:theta}) of the form $\theta_s=2\pi r/M$, where $M$ divides $N$ and $0<r<M/2$ is an integer. Writing $N=qM$ with $q$ an integer, then $N\theta_s=2\pi rq$ and $\cos (N\theta_s)=1$, $\sin(N\theta_s)=0$, and $\sin(N\theta_s/2)=0$. The first two terms of Eq.~(\ref{eq:theta}) vanish, which implies that the third term must be zero for $\theta_s$ to be a solution. We distinguish two cases when this occurs: either $k'-k$, or equivalently $N-k'+k$, is divided by $M$; or $2(k'-k)$ is divided by $M$. 
These conditions define the singular states $\theta_s$.

To illustrate the occurrence of singular states, we consider $N=6$ as an example, which can be factorized by $M=2,3,6$. First, for $M=2$, there is no $r$ allowed, and hence no singular states associated to this value of $M$. For $M=3$ we have $r=1$ and $M$ divides $k'-k=3$, which defines the singular state $\theta_s=2\pi/3$. Finally, for $M=6$ we can have $r=1,2$, and $M$ divides $2(k'-k)=6$. The case $r=2$ yields the same singular state as $M=3$, while $r=1$ defines the singular state $\theta_s=\pi/6$. Then, for a $\mathcal{PT}$-symmetric ring of size $N=6$, and having the separation $k'-k=3$ between the gain and loss, the singular states correspond to the energies $E=\pm 1$. This case is illustrated in Fig.~\ref{fig:singular} (top) for $a=0.5$.

The reasoning used above is independent of $a$ and therefore it holds for $a=0$ too. In this case the eigenvalues $E=\pm 1$ are doubly degenerate for all $\eta$; see Fig.~\ref{fig:singular} (bottom). In this configuration the exceptional point observed at $\eta= 2$ is related to the eventual coalescence of the eigenvalues $E=\pm 2$ for $\eta=0$. Note that the same situation for $a=0$ is encountered if $N$ is even and $k'-k=N/2$: all singular states are real and doubly degenerated for $\eta\in[-2,2]$, where a $\mathcal{PT}$-symmetric phase exists. Indeed, from Eq.~(\ref{eq:theta}) with $a=0$, $N$ even and $k'-k=N-k'+k=N/2$, it follows that the non-degenerate eigenvalues satisfy
\begin{equation}
    \eta^2 + 4\cos^2\theta = \eta^2 + E^2 = 4,
    \label{eq:circle}
\end{equation}
which is a circle in the $E$--$\eta$ plane of radius 2 for $\eta^2 \leq 2$, or a hyperbola for imaginary $E$; cf. Fig.~\ref{fig:singular} (bottom). We emphasize that the $\mathcal{PT}$-symmetry phase described has zero onsite energies for all sites.

Finally, if $a\neq 0$, a tedious but simple calculation shows that for the singular states described thus far we have $u_k= u_{k'}=0$. This implies that these states do not couple with the source and sink, and cannot be used for transport through the system. In \cite{OrtegaJPA2020} we called these kind of states ``opaque''.

\subsection{Accidental singular eigenstates}

\begin{figure}
  \centering
  \includegraphics[scale=0.45]{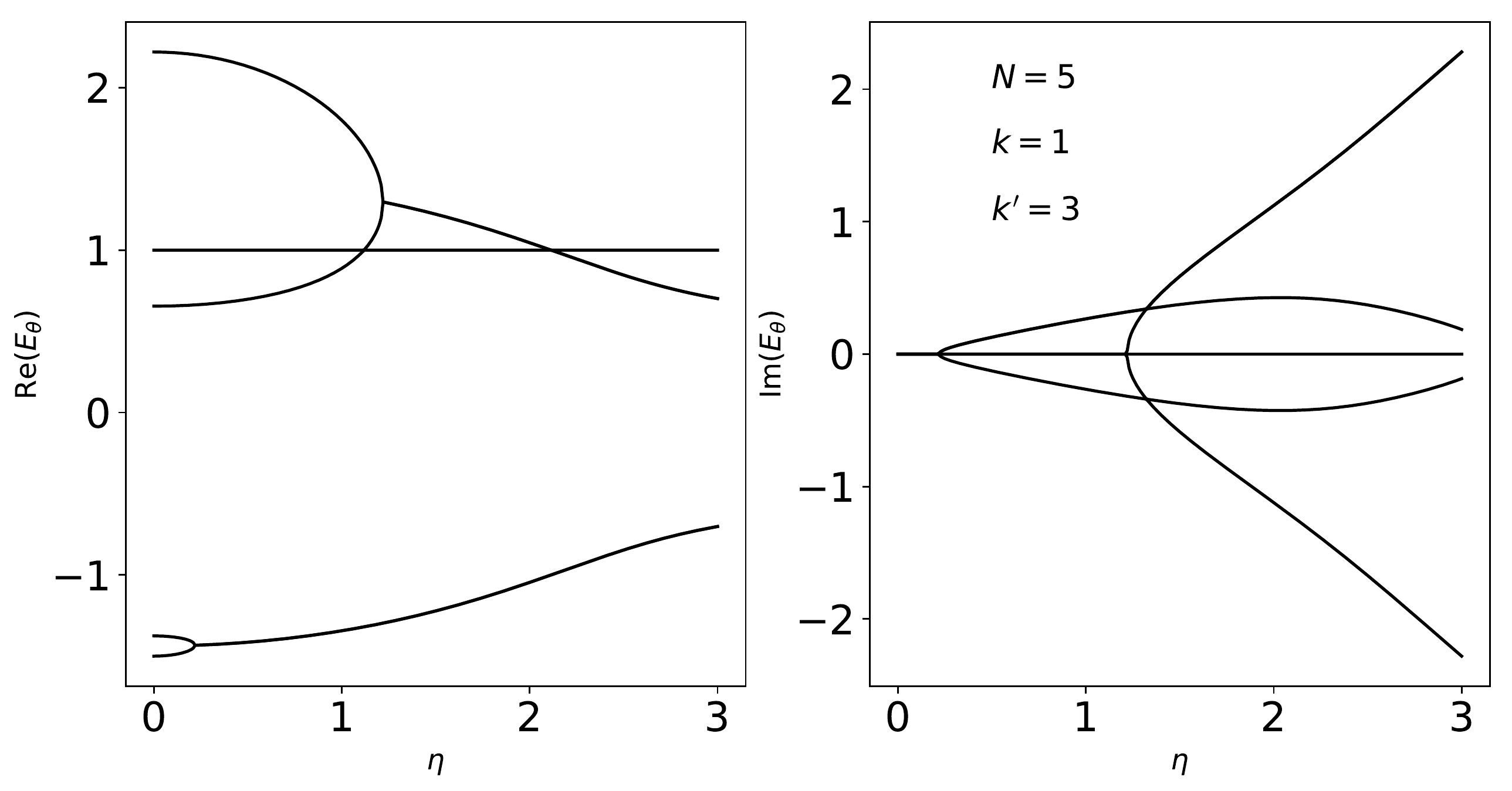}
  \caption{Spectrum of a $\mathcal{PT}$-symmetric ring 
  with an accidental singular state at $E=1$ ($N=5$, 
  $k'-k=2$, $a=0.5$.
}
  \label{fig:accidental}
\end{figure}

The introduction of the onsite energy at the contacts gives rise to a new kind of singular states that we refer to as ``accidental singular states''. These states correspond to values $\theta_a=\pi m/(k'-k)$ for $0<m<(k'-k)$ 
where we assume that $k'-k$ does not divide $N$. Similarly, the states $\theta_a=\pi m/(N-k'+k)$ also define accidental singular states, 
with $0<m<(N-k'+k)$ integer and with $N-k'+k$ not dividing $N$. With these definitions Eq.~(\ref{eq:accidental-states}) holds, i.e., the last term in Eq.~(\ref{eq:theta}) vanishes but not the first two. Yet, we can now choose the value of the onsite energy $a$ so that both terms cancel; specifically
\begin{equation}
    a=-\sin\theta_a \tan\left(N\theta_a/2\right).
    \label{eq:atune}
\end{equation}
An instance of an accidental singular state is illustrated in  Fig.~\ref{fig:accidental} for $N=5$ with $k'-k=2$ or $N-k'+k=3$. In this case we have accidental eigenstates when $\theta_a=\pi/3$, if $a=0.5$, corresponding to an energy $E=1$. Other accidental eigenstates in the system occur at $\theta_a=2\pi/3$ with $a=1.5$, and $\theta_a=\pi/2$ with $a=-1$. In contrast to the singular states discussed above, the amplitude of the eigenvectors does not vanish at the contacts. This situation corresponds to the states we called ``transparent'' in \cite{OrtegaJPA2020}, through which transport is efficient, independently of the value of $\eta$. 

\section{Eigenvalue reconversion and directional eigenstates}
\label{sec:egreconv}

We discuss now an interesting property of the system, that we term eigenvalue reconversion. As we have shown above, for small enough $\eta$ (and $a$ distinct from zero) we have a $\mathcal{PT}$--symmetric phase, which eventually is lost through an eigenvalue coalescence at an EP. In addition, for $\eta\to\infty$ we showed that two eigenvalues have an asymptotically increasing imaginary part, whereas in the same limit, the $N-2$ remaining eigenvalues have imaginary parts that either vanish as $\eta$ increases, or are identically zero. Interestingly, this latter case may occur via the reconversion of a pair complex conjugate eigenvalues back to being purely real again at a reverse EP.

\begin{figure}
  \centering
  \includegraphics[scale=0.5]{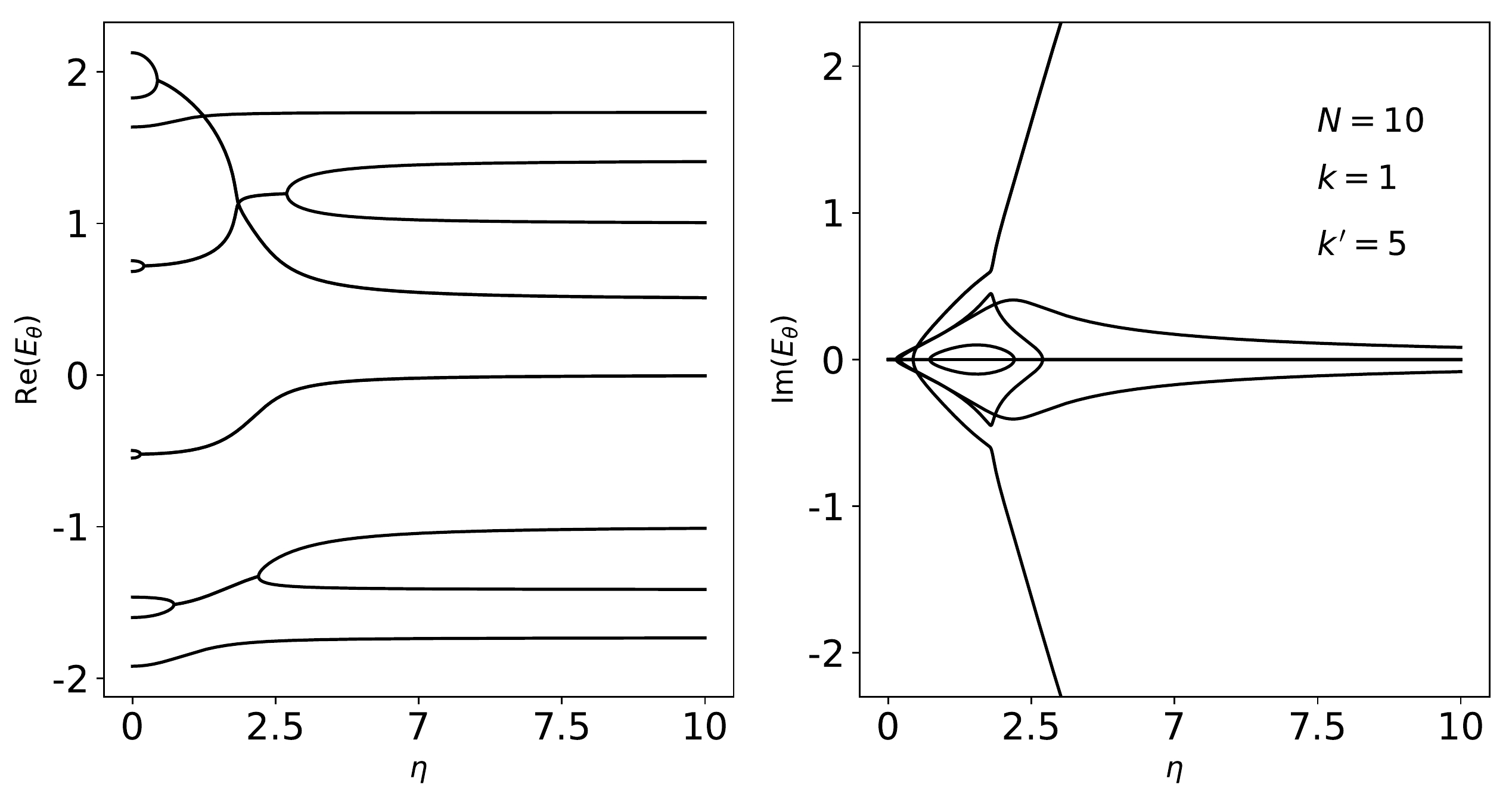}
  \caption{Real and imaginary parts of the spectrum in terms of $\eta$, illustrating the eigenvalue reconversion, for $N=10$, with the gain and loss located at $k=1$ and $k^\prime=5$, for $a=0.5$. Note that there are two pairs of eigenvalues that after a second coalescence they undergo a transition from complex back to purely real.}
  \label{fig:5}
\end{figure}

The phenomenon of eigenvalue reconversion is illustrated in Fig.~\ref{fig:5}, which shows the case $N=10$, with $k=1$, $k'=5$ and $a=0.5$. It is apparent that two energies have imaginary parts that diverge to $\pm \infty$ for $\eta$ large, and two eigenvalues remain real for all values of $\eta$, though they are not singular states since they have a weak dependence on $\eta$. The remaining six eigenvalues became complex after coalescences at EPs, four of them become real again after a new coalesces at reverse EPs. We notice that the two eigenvalues that still have imaginary parts that tend asymptotically to zero as $\eta$ grows, correspond to $\theta_0=\pi/2$; thus, using Eq.~(\ref{eq:theta1OddCase}) for $\theta_1$, yields the quasi-momenta
\begin{equation}
    \theta \sim \frac{\pi}{2}\pm \frac{i}{\sqrt{6}|\alpha|},
    \label{eq:Ea}
\end{equation}
indicating imaginary parts indeed vanish slowly for this case, as opposed to what happens at the reverse EPs. 

\begin{figure}
  \centering
  \hspace*{-5mm}
  \includegraphics[scale=0.37]{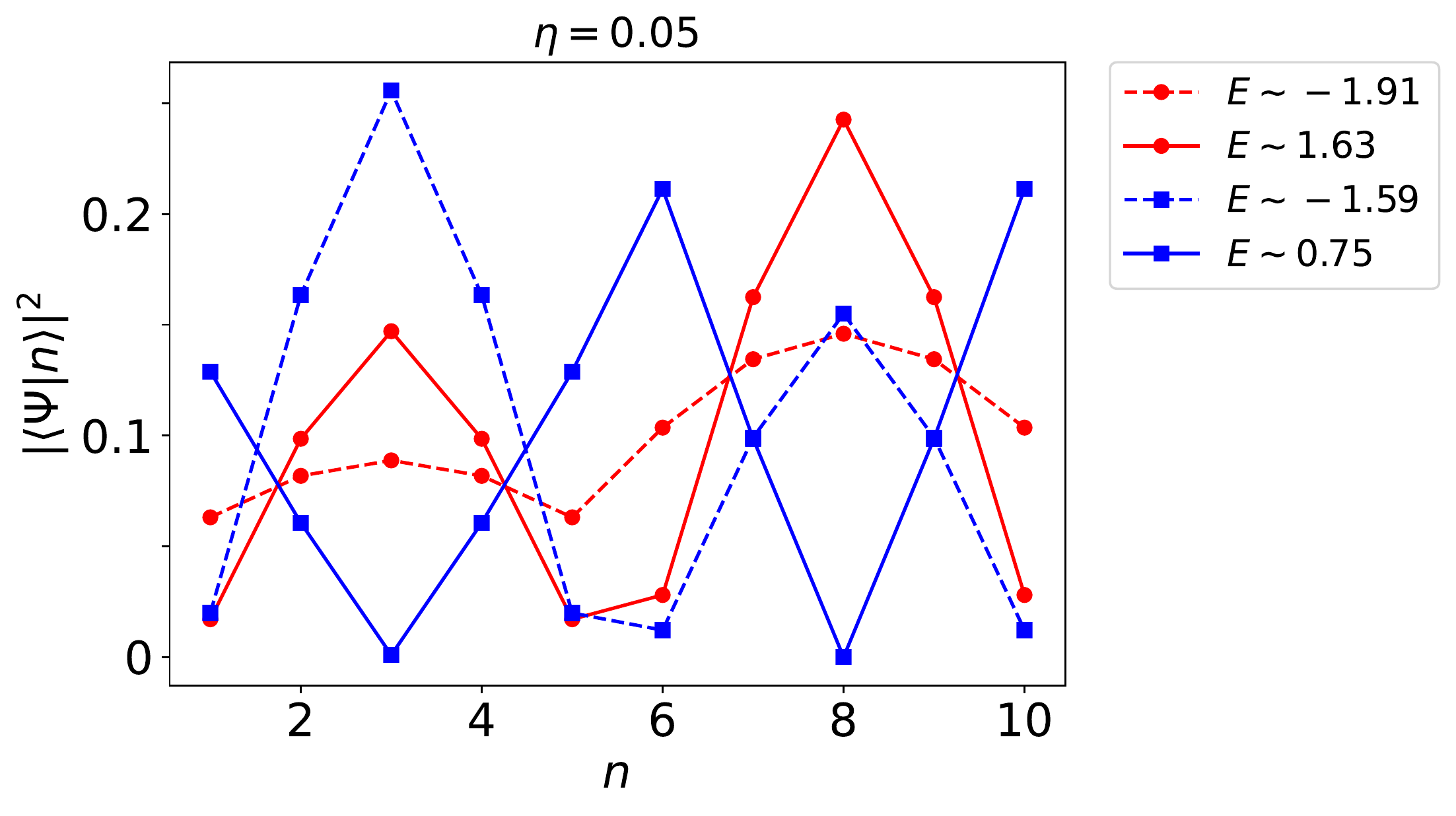}
  \includegraphics[scale=0.37]{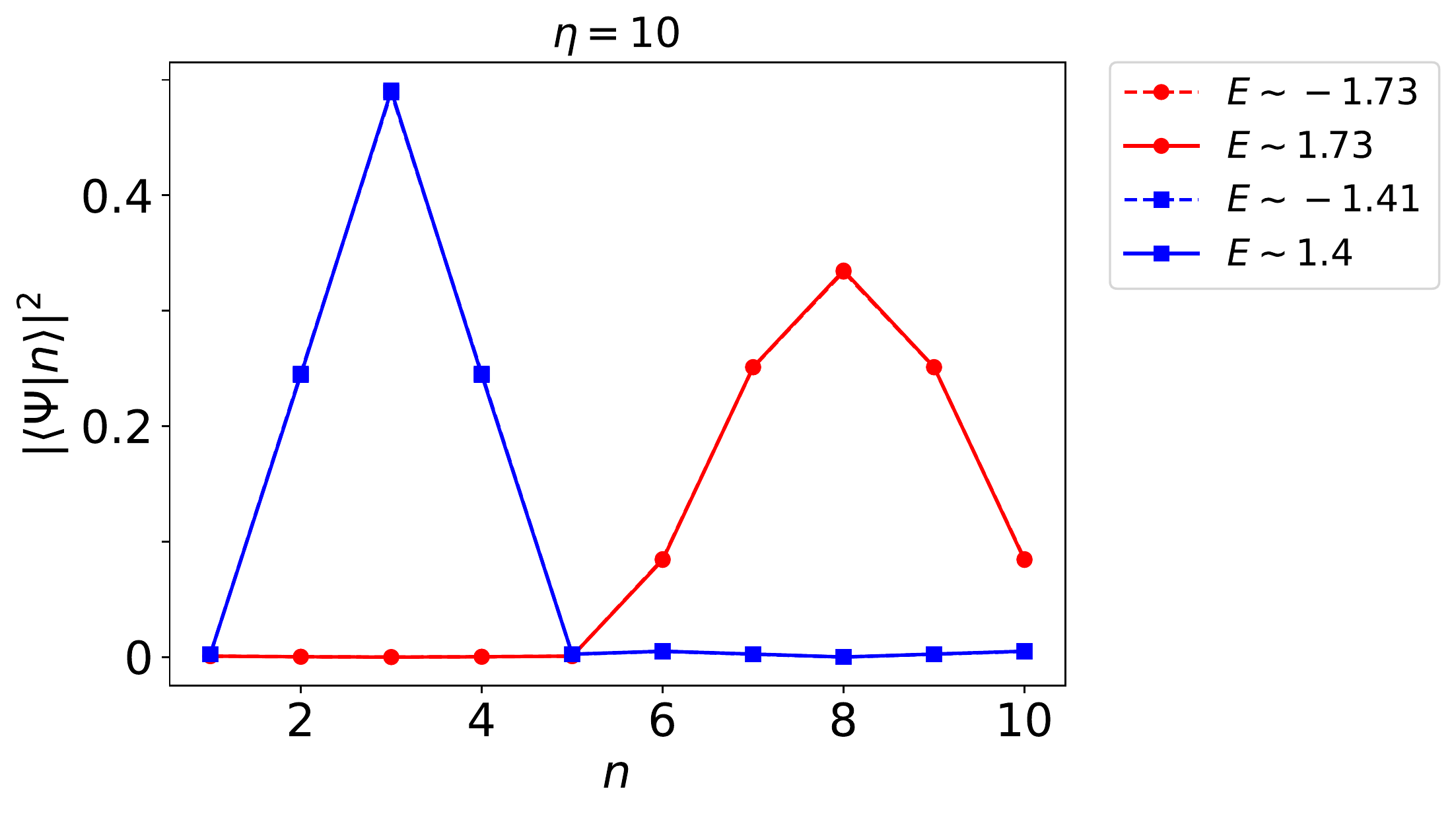}
  \caption{Eigenvectors in the site basis for some real eigenvalues illustrated in Fig.~\ref{fig:5}. These eigenvectors are not singular because the associated energies depend on $\eta$. Eigenvectors in red correspond to eigenvalues that are real for all $\eta$; eigenvectors in blue are related to
  energies that 
  undergo a reconversion. Note that for $\eta=10$, each eigenvector is localized to the left or right of the chain.}
  \label{fig:6}
\end{figure}

As described in Sect.~\ref{sec:ModelSolution}, for large values of $\eta$ the spectrum effectively splits in four parts: two levels are associated with localized states at the gain and loss positions, and then we have two separated open chains between the contacts whose eigenvalues are asymptotically given by Eq.~(\ref{eq:accidental-states}). According to this effective decoupling, we expect that some states are non-zero in one of the sub-chains between the contacts, and vanishing small in the other. In this case, whatever transport that may happen from the gain to the loss, can only occur along one branch of the ring. We call these states ``directional eigenstates''. 
Figure~\ref{fig:6} illustrates the modulus squared for a couple of these eigenvectors in the site basis; the parameters used for the ring correspond to those used in Fig.~\ref{fig:5}. The eigenvectors in red correspond to energies that remain real for all $\eta$; the eigenvectors in blue correspond to a couple of energies that suffer a reconversion. The specific values $\eta$ and $E$ are given in the figure. For small values of $\eta$, cf. the left panel in Fig.~\ref{fig:6}, the eigenvectors are more or less evenly distributed throughout the chain. 
The directional eigenstates become clear when $\eta$ is large compared to the value when the reconversions occur; see Figure~\ref{fig:6}, right panel. We note that both eigenvectors in red tend to the same local probability $|\langle \Psi|n\rangle|^2$ and are localized on the right of the chain; the same statement holds for the eigenvectors illustrated in blue, which localize on the left of the chain. 

\begin{figure}
  \centering
  \includegraphics[scale=0.37]{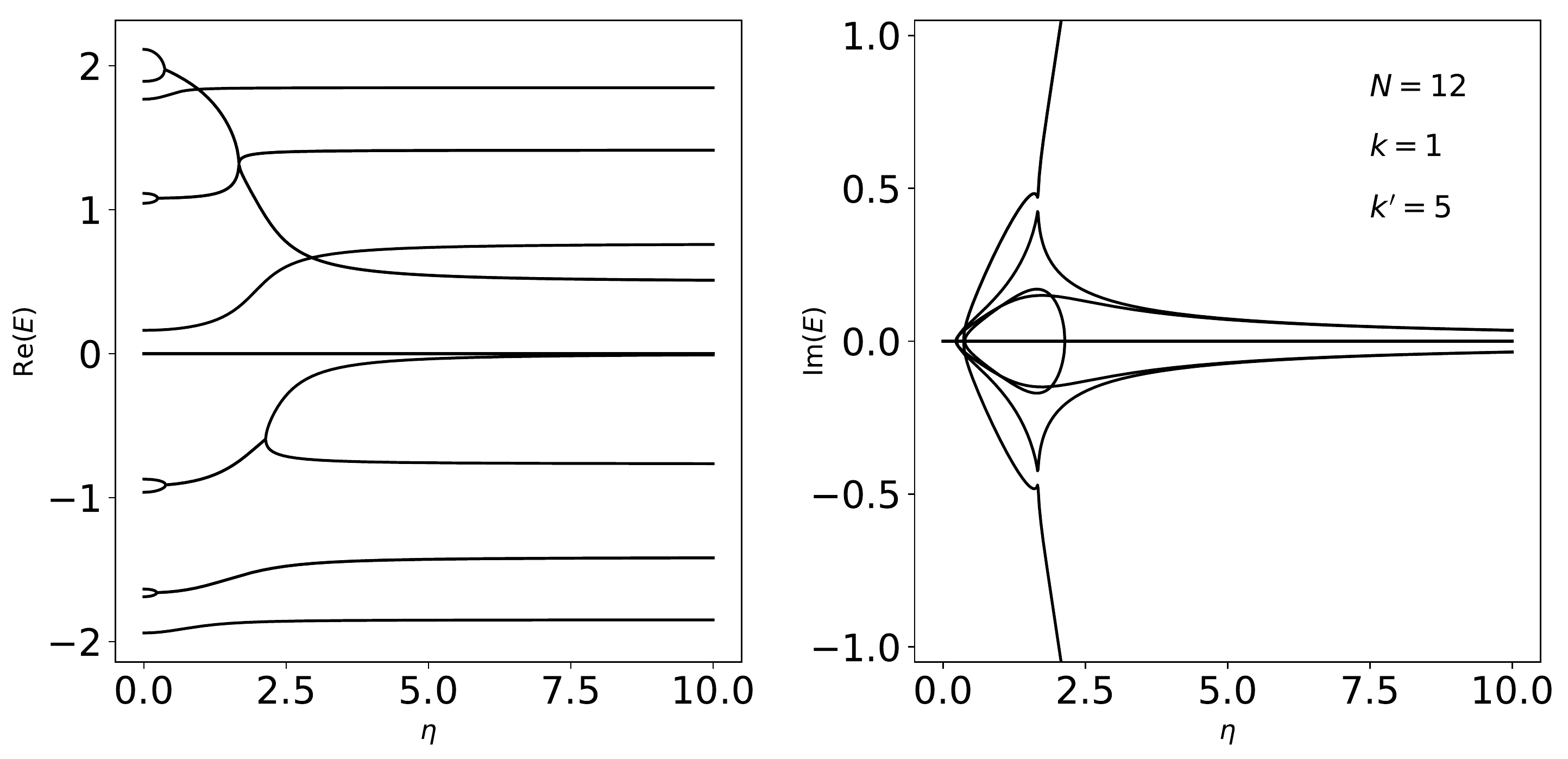}
  \hspace*{-5mm}
  \includegraphics[scale=0.37]{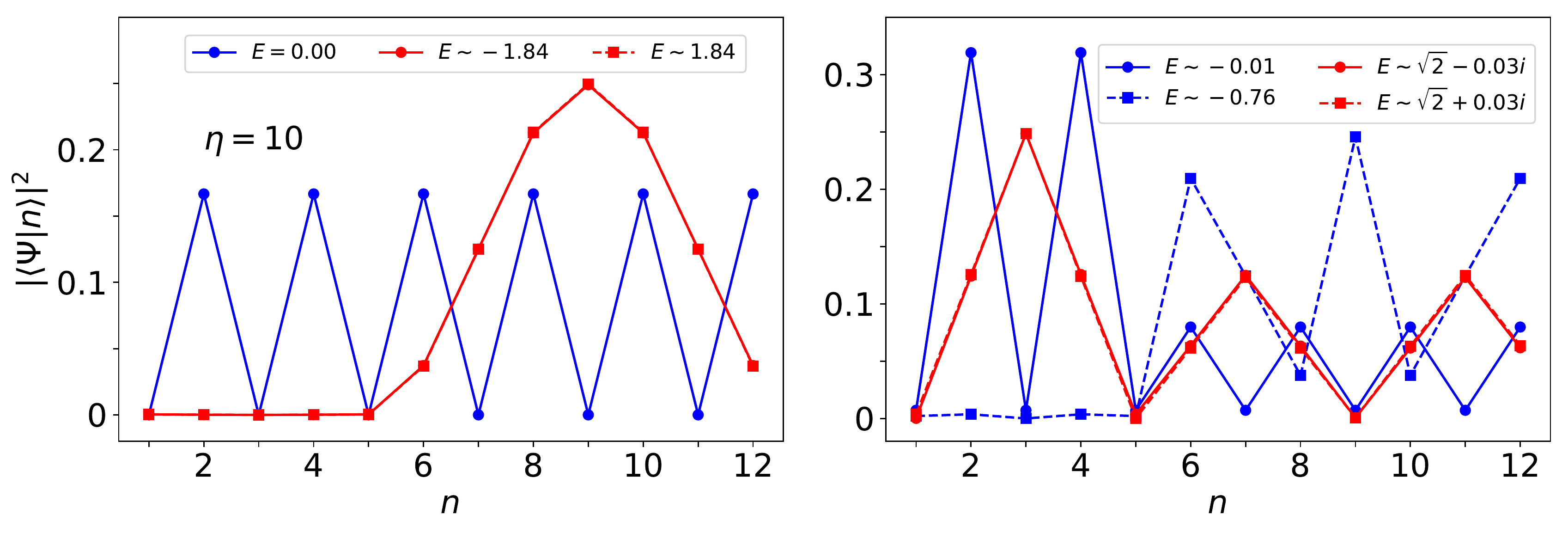}
  \caption{Similar as Fig.~\ref{fig:6} but for $N=12$, $k=1$, $k'=5$ and $a=0.5$. Top
  row: spectrum as a function of $\eta$. Bottom row: Modulus square of several eigenvectors. On the left, the eigenvector in blue is opaque (its energy does not vary with $\eta$); in red two eigenvectors whose eigenvalues remain real but vary with $\eta$. In this last case, the eigenvectors are localized in the long branch of the ring. On the right, the red eigenstates correspond to energies whose imaginary part tend to zero as $\eta$ grows, while the blue ones to energies that have been reconverted and are not localized in either branch of the ring. Note that in this case, eigenvalue reconversion does not imply that both eigenvectors tend to the same shape, c.f. Fig.~\ref{fig:6} right panel.}
  \label{fig:7}
\end{figure}

This eigenvector localization can be understood qualitatively by recalling that as $\eta$ increases, the branches of the ring become effectively decoupled. In this regime, we can consider the eigenstates and eigenvalues of each branch separately. For each eigenvalue that is not shared between both branches, an eigenstate of the complete system can be constructed by considering a vector that coincides with the corresponding eigenstate of the appropriate branch, and vanishes on the other branch, as is observed. 
If both branches share an eigenvalue, the corresponding eigenstates do not necessarily localize in one branch as discussed. This situation is exemplified for a ring with $N=12$, $k=1$, $k^\prime=5$ and $a=0.5$ in Fig.~\ref{fig:7}. In this case, in the large $\eta$ limit, the ring is split in two branches with 3 and 7 sites, aside from the leads. Following~\cite{OrtegaJPA2020}, the quasi-momenta that yield the energies of each branch read $\theta_{\rm short}=\pi r/4$, $r=1,\dots, 4$, and $\theta_{\rm long}=\pi s/8$, with $s=1, \dots, 8$. Clearly, all quasi-momenta from the short branch have a corresponding quasi-momenta in the long branch, and thus the eigenvalues in the short branch are eigenvalues of the long one; this indicates that no eigenvector will be localized in the short branch. This is illustrated in Fig.~\ref{fig:7}: The top panel shows the energy spectrum, where we can see that two pairs of eigenvalues have an imaginary part that asymptotically tends to zero; the bottom panels illustrate some of the eigenstates of the system. The red eigenstates in the bottom left panel correspond to eigenvalues that are real for all $\eta$, and the blue one is a singular state shared by both branches. In the bottom right panel, two of the states illustrated are not localized in either branch; the dotted blue state localizes on the right branch. In this panel, the red states correspond to the eigenvalues whose imaginary parts asymptotically vanish, and the blue ones to real eigenvalues that have undergone a reconversion. 

\section{Summary and Conclusions}
\label{sec:Conclusions}

In this article we have studied the $\mathcal{PT}$-symmetric ring with on-site potentials $a$, in addition to the $\pm i\eta$ representing the strength of the input and output at the gain and loss sites. 
We discussed the conditions under which the system possesses an unbroken $\mathcal{PT}$-symmetric phase. If there is no such phase, as happens in general when there is only gain and loss ($a=0$), the eigenvalues are complex and the associated singularities in the complex plane correspond to diabolical points. 
The introduction of a real onsite potential allows the system to present a $\mathcal{PT}$ unbroken phase for a range of values of the parameter $\eta$, until exceptional points are encountered.
We addressed the transport properties of the system for non-degenerate $\mathcal{PT}$-symmetric eigenstates. We showed that the system presents opaque as well as transparent singular eigenstates; the latter only appear at precise values of the onsite potential $a$. We termed such states ``accidental singular states'', to distinguish them from the former set, that also appears for the one-dimensional chain with open boundary conditions~\cite{OrtegaJPA2020}. 

The study of the local currents reveals that for certain $\mathcal{PT}$-symmetric eigenstates, a stationary back-flow (from the sink to the source) is possible along one of the branches of the ring. We also observed another interesting property of the eigenvalues of the ring, the eigenvalue reconversion, in which pairs of eigenvalues, after becoming complex at an EP, under certain circumstances undergo a reverse EP and become real again as $\eta$ increases. 
Finally, we found that in the large $\eta$ regime, the eigenstates may show a partial localization effect, becoming localized in only one branch of the ring. In this situation, any transport through the system may only occur in one of the branches, becoming, in some sense, ``directional eigenstates''.

The analysis presented thus far has revealed that the periodic boundary conditions allow for an unexpectedly rich behaviour of the eigenvalues and eigenvectors of the system. From the transport point of view, one can tune the parameters of the system to modify the behavior of the system. 
In addition, the system can have a backflow of density from the sink to the source, which also gives rise to a effective circulating current around the ring. We plan to study the implications of this circulating current in a future work.

Finally, the number of sites needed to build systems that display the behaviors discussed in this paper fits well with the current capabilities of microwave experiments~\cite{DietzPS2019,StegmannPRB2020}. We believe our results can be observed in such platforms.

\begin{acknowledgments}
We gratefully acknowledge financial support from the UNAM-PAPIIT research grant IG-100819. AO is grateful for the support of the National Research, Development and Innovation Office of Hungary (Project K124351) and the Quantum Information National Laboratory of Hungary.
\end{acknowledgments}

\appendix

\section{Solution of the eigenvalue problem}
\label{sec:egvalsol}

In this section we present a simple variation of the method of Losonczi-Yueh~\cite{Losonczi1992, Yueh2005}
to obtain the eigenvalues and eigenvectors of the tight-binding
Hamiltonian with periodic boundary conditions.

\begin{equation}
  H = \sum_{i=1}^{N-1}\big(\ket{i}\bra{i+1} + \ket{i+1}\bra{i} \big)
  + \ket{1}\bra{N} + \ket{N}\bra{1} + \big(\alpha\ket{k}\bra{k} + \beta\ket{k'}\bra{k'} \big),
  \label{eq:hamapp}
\end{equation}
with $\alpha, \beta\in \mathbb{C}$ and $k<k'$, the positions of the
contacts, arbitrary. The method is as an alternative
to the popular Bethe ansatz\cite{ScottPRA2012}, though from our point of
view, our method is more transparent. 

The solution for a one dimensional tight-binding chain with
open boundary conditions has been computed in~\cite{OrtegaJPA2020}.
We follow that derivation here, 
and only highlight the main differences due to the periodic
boundary conditions. Note that the Hamiltonian~(\ref{eq:hamapp})
is more general than Eq.~(\ref{eq:ham}), and it does not have the
$\mathcal{PT}$ symmetry. Clearly, $\mathcal{PT}$ symmetry is obtained
by setting $\beta=\alpha^*$.

Let $E$ be an eigenvalues of $H$ and $\ket{u} = \sum_j u_j \ket{j}$ 
its associated eigenvector. The eigenvalue problem $H u = Eu$ can be written 
as a set of linear equations

\begin{equation}
   \begin{split}
     u_0 &= u_N,\\
     u_0 + u_1 + u_2 &= E u_1,\\
     \vdots \\
     u_{k-1} + u_k + u_{k+1} &= (E - \alpha) u_k,\\
     \vdots \\
     u_{k'-1} + u_{k'} + u_{k'+1} &= (E - \beta) u_{k'},\\
     \vdots \\
     u_{N-1} + u_N + u_{N+1} &= E u_N, \\
     u_{N+1} &= u_1,
   \end{split}
   \label{eq:Hsyseq}
\end{equation}
where $u_0 = u_N$ and $u_{N+1}=u_1$ correspond to the periodic boundary 
conditions. The idea of the Losonczi-Yueh method is to rewrite the system of 
equations as an equation for infinite sequences; a complete introduction 
to the method can be found in~\cite{Chengbook2003}. Following 
Yueh~\cite{Yueh2005}, one substitutes
$E=2\cos\theta$ and after some algebra with infinite sequences one finds
that the infinite sequence $u = \lbrace u_0,u_1,\dots,u_{N+1},0,\dots\rbrace$ 
has components
\begin{equation}
   \begin{split}
     u_j &=\frac{1}{\sin\theta}\Big[ u_0\sin[(1-j)\theta] + 
     u_1\sin(j\theta) - \alpha u_k\sin[(j-k)\theta] 
     \Theta(j-k-1)\\ 
     & \qquad - \beta u_{k'}\sin[(j-k')\theta]\Theta(j-k'-1) \Big],
   \end{split}
   \label{eq:ujapp}
\end{equation}

where $\Theta (n)=0$ for $n<0$ and $\Theta (n)=1$ for $n\ge 0$. The 
values of $u_k$ and $u_k'$ can be found from this equation by 
setting  $j=k,k'$; they read
\begin{equation}
   \begin{split}
     u_k &= \frac{1}{\sin\theta} \Big[
        u_0\sin[(1-k)\theta] + u_1\sin(k\theta) \Big],\\
     u_{k'} &= \frac{1}{\sin\theta} \Big[ 
        u_0\sin[(1-k')\theta] + u_1\sin(k'\theta)\\ 
     & \qquad - \alpha \frac{\sin[(k'-k)\theta]}{\sin\theta} 
       \big(u_0\sin[(1-k)\theta] + u_1\sin(k\theta)\big)\Big].
   \end{split}
   \label{eq:ukl}
\end{equation}

From Eqs.~(\ref{eq:ujapp}) and (\ref{eq:ukl}) one can see that 
the $u_j$ component depends on $u_0$ and $u_1$. In order to find their 
values, we exploit the periodic boundary conditions, $u_N=u_0$ and 
$u_{N+1} = u_1$. 
The result can be written as an homogeneous linear system of equations
\begin{equation}
\begin{pmatrix}
\mathfrak{a}(\theta) & ~\mathfrak{b}(\theta) \\
\mathfrak{c}(\theta) & ~\mathfrak{d}(\theta)
\end{pmatrix}
\begin{pmatrix}
u_0 \\ u_1
\end{pmatrix}
= 0,
\label{eq:systemu0u1}
\end{equation}
where each of the matrix components 
are given explicitly as
\begin{eqnarray}
    \mathfrak{a}(\theta) &=& \frac{1}{\sin\theta}\bigg[ \sin\theta+\sin[(N-1)\theta]
    -\alpha\frac{\sin[(N-k)\theta]\sin[(k-1)\theta]}{\sin\theta}
    \nonumber\\
    && -\beta\frac{\sin[(N-k')\theta] \sin[(k'-1)\theta]}{\sin\theta}
    + \alpha\beta \frac{\sin[(N-k')\theta] \sin[(k'-k)\theta] \sin[(k-1)\theta]}{\sin^2\theta} \bigg],
    \\ \label{eq:a}
    \mathfrak{b}(\theta) &=& -\frac{1}{\sin\theta} \bigg[\sin(N\theta)
    - \alpha\frac{\sin[(N-k)\theta] \sin(k\theta)}{\sin\theta} \nonumber\\
    && - \beta\frac{\sin[(N-k')\theta]\sin(k'\theta)}{\sin\theta}
    + \alpha\beta\frac{\sin[(N-k')\theta]\sin[(k'-k)\theta] \sin(k\theta)}{\sin^2\theta}  \bigg],
    \\ \label{eq:b}
    \mathfrak{c}(\theta) &=& \frac{1}{\sin\theta} \bigg[\sin(N\theta)
    - \alpha\frac{\sin[(N-k+1)\theta] \sin[(k-1)\theta]}{\sin\theta} \nonumber\\ 
    && - \beta\frac{\sin[(N-k'+1)\theta]\sin[(k'-1)\theta]}{\sin\theta}
    +\alpha\beta\frac{\sin[(N-k'+1)\theta]\sin[(k'-k)\theta] \sin[(k-1)\theta]}{\sin^2\theta} \bigg],\nonumber\\
    \\
    \label{eq:c}
    \mathfrak{d}(\theta) &=& \frac{1}{\sin\theta}\bigg[ \sin\theta-\sin[(N+1)\theta]
    +\alpha\frac{\sin[(N-k+1)\theta] \sin(k\theta)}{\sin\theta}\nonumber\\
    &&+\beta\frac{\sin[(N-k'+1)\theta] \sin(k'\theta)}{\sin\theta}
    -\alpha\beta\frac{\sin[(N-k'+1)\theta] \sin[(k'-k)\theta] \sin(k\theta)}{\sin^2\theta} \bigg].
    \label{eq:d}
\end{eqnarray}
%

This set of equations 
has a 
non-trivial solution when the corresponding determinant of the system 
vanishes; this defines the equation for the quasi-momentum $\theta$.
After a trivial, albeit long, algebra the equation for $\theta$ reads
\begin{equation}
   4\sin^2\left(\frac{N\theta}{2}\right) + 
   \frac{\alpha+\beta}{\sin\theta} \sin(N\theta) - 
   \frac{\alpha\beta}{\sin^2\theta} \sin[(k'-k)\theta] 
   \sin[(N-k'+k)\theta]= 0.
   \label{eq:thetaapp}
\end{equation}

Once the condition (\ref{eq:thetaapp}) is satisfied, the eigenvectors 
are obtained using (\ref{eq:ujapp}) and (\ref{eq:ukl}), 
by writing $u_1$ in terms of $u_0$ from one of the equations 
(\ref{eq:systemu0u1}); $u_0$ is fixed by imposing a normalization for 
the eigenvectors.

\bibliographystyle{unsrt}
\bibliography{bibliography.bib}

\end{document}